\documentclass[12pt,a4paper]{article}

\usepackage{amsmath,amsfonts,latexsym}

\usepackage{graphicx}

\newcommand{\Z}{\mathbb Z}

\newcommand{\R}{\mathbb R}

\newcommand{\xor}{\oplus}

\newenvironment{proof}{\textit{Proof:}}{\hfill $\Box$\medskip}

\newtheorem{prop}{Proposition}
\newtheorem{remark}[prop]{Remark}

\begin{document}

\title{An attack on MySQL's login protocol}
\author{Ivan~F.F.~Arce\and Emiliano Kargieman \and Gerardo Richarte \and Carlos 
Sarraute \and Ariel Waissbein}
\date{January 24, 2002}

\maketitle

\begin{abstract}
The MySQL challenge--and--response authentication protocol is proved 
insecure. We  show how can an eavesdropper impersonate a valid user 
after witnessing only a few executions of this protocol. The algorithm
of the underlying attack is presented. Finally we comment about
implementations and statistical results.
\end{abstract}

\begin{section}{Introduction}
The use of computer-based user authentication has become a cryptographic 
tool widely used in these days. Every remote connection (SSH, SSL, 
et~cetera) is initiated with a user authentication. This also holds 
true for remote access databases as the {\sc MySQL Database Engine}. 
Computer-based user authentication amounts to one of different process 
by which an entity, the user, is able to authenticate himself by way of 
a cryptographic protocol to another entity, often a server, in such a 
way that no other person ---than the user--- can do this. Different 
standards exist for user authentication such as zero--knowledge 
cryptography (e.g., Fiat--Shamir \cite{fiat87how}, Guillon--Quisquater 
\cite{quisquaterpractical} or Schnorr
\cite{schnorr89efficient,schnorr91efficient} identification protocols), or 
challenge--and--response protocols (e.g., see Kerberos \cite{kerberos},
the Secure Shell authentication protocols \cite{itef.ssh} see also 
\cite{TomWu}). 

{\sc MySQL Database Engine} (\cite{mysql}) is a world known open source 
database engine enabling remote access through secured channels. This 
package is widely used in many applications such as world wide web portals 
and intranet services and has become a standard in its category.

The MySQL scenario is constituted by a {\em server}, which centralizes all the 
information in a database to which validated users, called {\em clients}, have 
access by logging on the server through an authentication procedure. When a
client authenticates himself to the server, he can then start a session and 
succeedingly obtain any information from the database in the server. This 
information then, travels through information channels encrypted with a key 
negotiated between the client and server, and can thus, only be read by this
client.
Different parameters as to how this is done can and are selected by
server administrators. However, all this possible configurations have in 
common the same authentication procedure.

The authentication protocol is designed by the MySQL team with a twofold purpose, 
to prevent the flow of plaintext  passwords over the network, and the storage of 
them in plaintext format on the server's and user's respective terminals. 
For these purposes, a challenge--response mechanism for authentication is chosen
together with a hash function. 
There is no mention of this authentication mechanism in the literature
as it was designed by the MySQL development team and never published.
The authentication procedure we describe here is extracted from the source 
code\footnote{available at {\tt http://www.mysql.com}} implemented on every 
version of {\sc MySQL}. Slight variations are to be found between versions
prior to {\tt 3.20}, and versions after {\tt 3.21}.

Regrettably, this authentication mechanism is not cryptographically
strong. Firstly we shall see that the second objective is not met, 
since the only value needed to authenticate a user is stored
both in his machine as in the server. But moreover, we shall see
that, each time a user underpasses a challenge--and--response 
execution, information allowing an attacker to recover this user's 
password is leaked. 

In view of these vulnerabilities which we describe 
in Section \ref{attack} we designed an attack ---described in Subsection 
\ref{attack.algoritmo}--- which permits an eavesdropper to authenticate 
to the database engine impersonating the witnessed valid user after only
witnessing a few successful authentications of this user.

We shall prove that all the contents of a MySQL database can be obtained
by sniffing a few client authentications. In fact, our algorithmic construct 
works in such a way that, for every time a client authenticates himself to
the server he narrows the key search space almost in an exponential 
manner. That is, this is done in such a way that starting with a brute--force 
key search space of $2^{64}$, we are reduced to a search space of $300$ 
after only $10$ authentications!

Previous vulnerabilities in the {\sc MySQL Database Engine} where discovered
in the Bugtraq advisories \cite{bugtraq1}, \cite{bugtraq2} and \cite{bugtraq3}.
An advisory authored by the subscribers which briefly describes this attack 
appeared as a Bugtraq Advisory  in \cite{core}.

\end{section}
\begin{section}{Technical Description}
\label{attack}
\begin{subsection}{The challenge/response mechanism}

The authentication protocol of {\sc MySQL Database Engine} is of the 
challenge--and--response type, the underlying idea behind this construction is that no 
passwords are sent between client and server through the connection.
The challenge-response mechanism of {\sc MySQL} does the following
(From {\tt mysql-3.22.32/sql/password.c}).
{\sc MySQL} provides users with two primitives used for the authentication
protocol: 

\begin{itemize}
\item a hash function, and
\item a (supposedly) one--way function;
\end{itemize}
both of their own design. The protocol goes as follows. On connection, when a 
user wants to log in, a random string is generated by the server and sent to the
client ---this is the challenge. The client, using as input the hash value of 
the random string he has received and the hash value of his password, calculates 
a new string using the one--way function ---the response--- which is sent to
the server.

This {\em checksum} string is sent to the server, where it is 
compared with a string generated from the stored {\tt hash\_value}
of the password and the random string. The password is saved (in 
{\tt user.password}) by using the {\tt PASSWORD(~)} function in mysql.
If the server calculates the same string as the response, the user 
is authenticated.

\end{subsection}
\begin{subsection}{Problem Description}

The hash function provided by {\sc MySQL} outputs eight-bytes strings, 
this makes $2^{64}$ possibilities. Whereas the one--way function
outputs eight-bytes strings, but with only $2^{45}$ possibilities, 
having a fixed input size of $8$ bytes. From this we deduce that more 
than one hashed password produces the same (expected) response for a 
given challenge. That is, we shall show that for a given challenge, 
not only the original password gives the correct response, but a much 
larger collection of values ---standing for different hashed 
passwords--- also do (see Section \ref{stats}).  We wish to remark that 
only the one--way function shall be analyzed and proved insecure, 
whereas the hash function shall not be analyzed.

We also point out that the authentication mechanism of {\sc MySQL} does 
not require the password for a successful authentication, but the password's
hash value. Hence, to impersonate a user only the hash value of this 
user's password is needed, so that the hash function is of no interest
for this account.

To validate our claim, we explain why the hash value of the password 
can be efficiently calculated using only a few executions of the
challenge--and--response mechanism for the same user. More explicitly,
in the forth--coming section we exploit this weakness, and deduce an
attack much more efficient than brute--force attack can be carried out
in only a few hours on a personal computer (see Section \ref{stats}).
Explicitly, after gaining a positive number of pairs of challenge and
response, an eavesdropper is able to efficiently calculate the set of
values, standing for hashed passwords, that pass the intercepted
challenge and response pairs.

To do this, firstly we describe how does the {\sc MySQL} one--way function
work and proceed to analyze the scheme's security, and then describe 
the attack we devised. The actual algorithm for making this calculations 
will be described in the Section \ref{attack.algoritmo}. The algorithm we 
describe was implemented in Squeak Smalltalk (see \cite{squeak}) by 
co--authors Gerardo Richarte and Carlos Sarraute. All the empirical results 
herein provided are derived from that implementation and the figures 1, 2 
and 3 (in subsections \ref{unpoligono} and \ref{deados}) are  cut--and--pasted 
black and white screen images of this implementation.

Let $n := 2^{30}-1$ (here $n$ is the {\tt max\_value} used in 
{\tt randominit( )} and {\tt old\_randoninit( )} respectively).
Fix a user ${\cal U}$. And initiate a challenge and response. 
That is, suppose the server has sent a challenge to the user
${\cal U}$. The hash value of this user's password is 8 bytes 
long. Denote by $p_1$ the first (leftmost) 4 bytes of this 
hash value and by $p_2$ the last 4 bytes (rightmost). Likewise,
let $c_1$ denote the first 4 bytes of the challenge's hash 
value and $c_2$ the last 4. We describe how to calculate the
output of the one--way function and how is this une--way function
used. 

\begin{itemize}
\item[{\bf 1}.]
calculate the values $s_1 := p_1\xor c_1$ and $s_2 := p_2\xor c_2$
(here $\xor$ denotes the bitwise exclusive or (X--or) function,
and $s_1$ and $s_2$ are the input to the one--way function),
\item[{\bf 2}.]
calculate recursively for $1\le i\le 8$:
$$
\begin{array}{rll}
	\begin{array}{r}
		s_1 \quad = 
	\end{array}
	&
	\begin{array}{l}
		s_1+3\cdot s_2 
	\end{array}
	&
	\begin{array}{l}
		\mbox{modulo } (n)
	\end{array} 
	\\ \\
	\begin{array}{r}
		s_2 \quad = 
	\end{array}
	&
	\begin{array}{l}
		s_1+s_2+33  
	\end{array}
	&
	\begin{array}{l}
		\mbox{modulo } (n)
	\end{array} 
	\\ \\
	\begin{array}{r}
		w_i \quad = 
	\end{array} 
	&
	\begin{array}{l}
		\displaystyle \left\lfloor \frac{31\cdot s_1}{n} \right\rfloor +64
	\end{array} 
\end{array} 
$$

\noindent (here $\lfloor x\rfloor:=\max\{k\in\Z:k\le x\}$ is the floor 
function)
\item[{\bf 3}.]
calculate form the preceding values
$$
\begin{array}{rll}
	\begin{array}{r}
		s_1 \quad = 
	\end{array}
	&
	\begin{array}{l}
		s_1+3\cdot s_2 
	\end{array} 
	&
	\begin{array}{l}
		\mbox{modulo } (n)
	\end{array} 
	\\ \\
	\begin{array}{r}
		s_2 \quad = 
	\end{array}
	&
	\begin{array}{l}
		s_1+s_2+33 
	\end{array} 
	&
	\begin{array}{l}
		\mbox{modulo } (n)
	\end{array} 
	\\ \\
	\begin{array}{r}
		w_9 \quad = 
	\end{array} 
	&
	\begin{array}{l}
		\displaystyle \left\lfloor \frac{31\cdot s_1}{n} \right\rfloor
	\end{array} 
\end{array} 
$$
\item[{\bf 4}.]
output the checksum value
$$
w=\Big( (w_1\xor w_9)\; \big\|\; \ldots\ldots\;\|\; (w_7\xor w_9)\; \|\; (w_9\xor w_9)
\Big)
$$
\end{itemize}

It is this checksum $w\in\{0,1\}^{64}$ that is sent, by ${\cal U}$, to the 
server. The server, that has in store the hash value of ${\cal U}$'s 
password, recalculates the checksum by this same process and 
succinctly verifies the authenticity of the value it has received. 
However it is a small collection of these checksums that allows any 
attacker to obtain $p_1$ and $p_2$ (the hash value of the user's 
password) and hence, enables the attacker to impersonate any user 
with only the information that travels on the wire between server and
client (the user ${\cal U}$). Actually, at each step of the algorithm
a set of values $(p_1,p_2)$ is calculated such that, 

The reason why the process of producing the checksum out of the hash
values of both the password and the challenge, is insecure is that 
this process can be efficiently reversed due to its rich arithmetic
properties. More specifically, consider the one--way function described 
above as a mapping $f$ that takes as input the two values $X$ and $Y$ 
and produces the checksum value $f(X,Y)=w$ (e.g., in our case 
$X:=p_1\xor c_1$ and $Y:=p_2\xor c_2$).
Then we can efficiently calculate all of the values $X^\prime,Y^\prime$ 
which map to the same checksum value than $X,Y$, i.e. if $f(X,Y)=w$, 
then we calculate the set of all the values $X^\prime,Y^\prime$ such
that $f(X^\prime,Y^\prime)=w$. This set is of negligible size in 
comparison to the $2^{64}$ points set of all the possible passwords' 
hashes in which it is contained. Furthermore, given a collection of 
challenge and response pairs made between the same user and the 
server, it is possible to efficiently calculate the set of all (hash values 
of) passwords passing the given tests.

\end{subsection}
\end{section}
\begin{section}{The Algorithm For the Attack}
\label{attack.algoritmo}
We now give a brief description of the attack we propose. This description
shall enable readers to verify our assertion that the {\sc MySQL} authentication
scheme leaks information. This attack has been implemented on Squeak 
Smalltalk and is now perfectly running. In what follows we shall depict the
procedures that constitute our attack. Since the attack is of a geometric nature,
we will be able to illustrate these procedures with screen snapshots of the
Squeak implementations.

The attack we designed is mainly divided into three stages. In these stages 
we respectively use one of our three algorithmic tools in various opportunities. 

\begin{itemize}
\item[\fbox{Procedure 1}] 
is an algorithmic process which has as input a checksum $w$, and outputs
a set of convex polygons ${\cal P}=\{P\}$ such that
  \begin{itemize}
	\item each $P\in{\cal P}$ is defined by its vertices,
	\item $f^{-1}(\{w\})=\bigcup_{P\in{\cal P}} P$, e.g. the point 
		$(p_1\xor c_1,p_2\xor c_2)$ of $\Z^2$ belongs to a polygon 
		$P$ in ${\cal P}$,
	\item $\#{\cal P}=36$ or $48$ (see Remark \ref{36o48} on the next
		subsection), and
	\item $\#(\Z^2\cap P)\sim 2^{33}$ (unproven estimate).
  \end{itemize}

\item[\fbox{Procedure 2}] 
An input is a pair of tuples $(c,{\cal P}),(c^\prime,{\cal P}^\prime)$ 
and an integer $k$, $1\le k\le 32$, where $c$ and $c^\prime$ are different 
pairs of challenges, with respective responses $w$ and $w^\prime$ 
such that $f^{-1}(w)=\cup_{\cal P}P$ and 
$f^{-1}(w^\prime)=\cup_{{\cal P}^\prime}P^\prime$, and 
such that ${\cal P}$ and ${\cal P}^\prime$ are compliant with the four conditions 
stated above. The output is a collection of polygons $\widetilde{\cal P}$ 
consisting of all the polygons $\widetilde P=\bigcup (P\cap Q)$, the union 
taken over all the squares $Q$ of side $2^{32-k}$ and vertices with entries in 
$\{i\cdot 2^k:0\le i\le 2^{32-k}\}$, and all the polygons $P\in{\cal P}$ in ${\cal P}$ 
for which there exists a $P^\prime\in{\cal P}^\prime$ such that 
$(P\xor c)\cap (P^\prime\xor c^\prime)\neq\emptyset$.

\item[\fbox{Procedure 3}]  
Given a set of integer points ${\cal Z}$ and a collection of pairs $(c^{(1)},w^{(1)}),
\ldots, \linebreak
(c^{(t)},w^{(t)})$ as input ($n\in\Z, n\ge 2$), this procedure outputs a new collection 
of points ${\cal Z}^\prime$ such that every point $z$ in ${\cal Z}^\prime$ passes the
$n$ given challenges (i.e. $f(z,c^{(k)})=w^{(k)}$ for all $z\in{\cal Z}^\prime$).
\end{itemize}

The rest of this section goes as follows, the three first forth--coming subsections
correspond to the three algorithmic tools we just described. On the fourth and last
subsection we explain how are these tools used in the attack and prove the attack
effective.

\begin{subsection}{From brute--force to brute forge}
In the preceding paragraphs of this section we have stated the input/output of 
Procedure 1. This subsection is devoted to this procedure we call algorithmic 
tool number 1, which we follow to describe. As explained, Procedure 1, from a 
pair of challenge and response $(c,w)$ produces a set of polygons ${\cal P}$ 
containing all the hashed passwords passing the same challenge, i.e. ${\cal P}$
is such that $f^{-1}(w)= \cup_{P\in{\cal P}}(P\xor c)$. We explain why are the 
elements of ${\cal P}$ convex polygons and why is it the case that this is a 
natural way of representing this particular set of points (i.e. $f^{-1}(w)$), 
subsequently we describe the algorithm underlying these procedure. 

For any (hashed) password $p$, and a (hashed) challenge $c$, the corresponding
response is calculated by the process described in the previous section. To invert 
this process, we inspect the one--way function $f$ more closely. 
Suppose with out loss of generality that $w_1,\ldots,w_8)$ are known (e.g. $w_9$ 
is known). Later in this section, in Remark \ref{w9}, we justify this supposition 
by explaining how is this problem tackled.

From the definition of the $w_1,\ldots w_8$ it follows that $64\le w_i<96$ (and for
$w_9$ it is $0\le w_9<32$). For the input $X=p_1\xor c_1$ and $Y=p_2\xor c_2$, it 
holds that the entries $w_i\in\Z$ verify a certain formula of the form 
$w_i=\lfloor \frac{31}{n}\bigl((\alpha_i\cdot X+\beta_i\cdot Y+\gamma_i\cdot 33)
\mbox{ mod }(n)\bigr)\rfloor+64$ for some integers $\alpha_i,\beta_i,\gamma_i\in\Z$, 
where the $\alpha_i,\beta_i,\gamma_i$ can be calculated for once and for all.
For example, 
$$
\begin{array}{l}
w_1 =\displaystyle\Big\lfloor \frac{31}{n}(3X+Y\mbox{ mod }(n))\Big\rfloor +64 ,\\ \\
w_2 =\displaystyle\Big\lfloor \frac{31}{n}(12X+5Y+33\mbox{ mod }(n))\Big\rfloor +64,\\ \\
\qquad\qquad\qquad\vdots \\ \\
w _8=\displaystyle\Big\lfloor \frac{31}{n}(322863X+140206Y+33\cdot 42450\mbox{ mod }
(n))\Big\rfloor +64,\\ \\
w_9=\displaystyle\Big\lfloor \frac{31}{n}(1389207X+603275Y+33\cdot 182656
\mbox{ mod }(n))\Big\rfloor .
\end{array}
$$

Notice that for the floor operation it holds
that $\lfloor x\rfloor\le x< \lfloor x\rfloor+1$. Then from the value of $w_1$, 
we deduce that the inequation
$$
\frac{n}{31}(w_1 -64) \le 3X+Y \mbox{ mod }(n)< \frac{n}{31}((w_1-64)+1)
$$
holds, i.e. there exists an integer $\delta_1\in\Z$ such that 
$\frac{n}{31}(w_1 -64)+\delta_1n \le 3X+Y< \frac{n}{31}((w_1-63)+\delta_1n$. 
Furthermore, form the fact that $0\le X,Y<2^{32}$ we deduce that 
$0\le \delta_1\le\lceil\frac{4\cdot 2^{32}}{n}\rceil -1=16$. 

For $w_2,\ldots,w_8$ similar equations hold. By a similar process to the one we
just applied to the defining equation of $w_1$, we deduce that
\begin{equation}
\label{eqs.}
\frac{n}{31}(w_i -64)+\delta_in \le \alpha_i\cdot X+\beta_iY +\gamma_i\cdot 33< 
\frac{n}{31}((w_i-63)+\delta_in
\end{equation}
where here the bounds for the $\delta_i$ can be analogously deduced, i.e., it follows 
that $\delta_i<\left\lceil\frac{2^{32}(\alpha_i+\beta_i)+33\gamma_i}{n}\right\rceil$. 

In this way, an attacker, for each choice of $\delta_1,\ldots,\delta_8$ is able
to construct the convex polygon
\begin{align*}
P_{\mathbf \delta}:=\displaystyle\bigcap_{1\le i\le 8} \Big\{(x,y)\in\R^2: \frac{n}{31}
(w_1 -64)+\delta_in\le \alpha_i\cdot X+\beta_iY +\gamma_i\cdot 33< \\
<\frac{n}{31}((w_1-63)+\delta_in\Big\},
\end{align*}
seen in Picture \ref{unpoligono}. We thus see that for every 
${\mathbf\delta}=(\delta_1,\ldots,\delta_8)$,  for
every integer point $(a,b)$ in $P_{\mathbf \delta}\cap\Z^2$ it holds that 
$f(c,(a,b))=w$. In fact it is easy to see that the other inclusion also holds, i.e., 
for every pair $(a,b)$ which is mapped via $f$ to the checksum $s$ there 
exist a tuple ${\mathbf\delta}=(\delta_1,\ldots,\delta_8)$ such that $(a,b)$ belongs 
to $P_{\mathbf \delta}$.

\begin{figure}
\centering
\includegraphics[width=7cm]{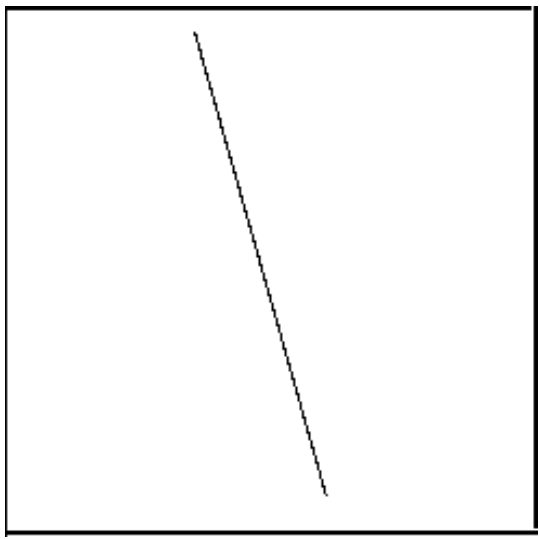}
\caption{A polygon}
\label{unpoligono}
\end{figure}

We shall see that many choices of the ${\mathbf\delta}$ will define the same polygon. 
Furthermore, we  prove that ${\cal P}:=\cup\, P$ (where the union is taken over all 
the possible tuples $\delta_1,\ldots,\delta_8\in\Z$), is a collection of $36$ or $48$   
polygons in the next remark.

\begin{remark}
\label{36o48}
Let ${\cal P}$ be defined as above. Then, each $P\in{\cal P}$ is a traslation of the other, i.e.
for every $P,P^\prime\in{\cal P}$ there exists $v\in\Z^2$ such that $P=P^\prime+v$.
Furthermore, it holds that $\#{\cal P}$ is equal to either $36$ or $48$.
\end{remark}
\begin{proof}
The traslation statement is straight--forward . What the procedure we described 
does, is constructing polygons by intersecting the convex polygons defined by
equations (\ref{eqs.}) and the square $[0,2^{32}]\times [0,2^{32}]$. The different 
polygons that appear in ${\cal P}$ are generated by the different choices of $k_i$
in the equations (\ref{eqs.}) and nothing else, this choices of $k_i$ shift the 
different defining lines $k_in$ upwards. The tangents of the different lines (which
define the polygons) respective to $w_1,\ldots,w_8$ are for every choice of 
challenge and password $3, 2.4, 2.3181,2.3052, 2.3031, 2.3028, 2.3027, 2.3027$ 
all rounded to the fourth decimal. 

To prove the second statement, let $P$ be a polygon in ${\cal P}$ with vertices 
$(a_1,b_1),\ldots,(a_t,b_t)$ it is easy to see that the polygon with vertices 
$(a_1+i\frac{n}{3},b_1+jn),\ldots,(a_n+i\frac{n}{3},b_n+jn)$ can be produced by a 
different choice of ${\mathbf \delta}$, for $i,j\in\Z$. Before proving this, we notice
that we are interested only in those polygons which have a nonempty intersection 
with the square $[0,2^{32}-1]\times [0,2^{32}-1]$ where the password is located. 
Since $\lfloor 2^{32}/\frac{n}{3}\rfloor =12$ and $\lfloor 2^{32}/ n \rfloor=4$, we 
deduce that $\#P\sim 48$.
\end{proof}

However, for our computational aims, we notice that the $\delta_i$ can be more 
accurately bounded by the formula re-studying the calculation process for the
$w_i$. The best bound for $\delta_1$ is the already calculated $\delta_1\le 16$. 
For $1\le i\le 8$ we write $s^{(i)}_1:=3s^{(i-1)}_1+s^{(i-1)}_2$ mod $(n)$, and 
$s^{(i)}_2:=s^{(i-1)}_1+s^{(i-1)}_2+33$ mod $(n)$ (where $s^{(0)}_1=X,s^{(0)}_2=Y$). 
For $1\le i\le 8$, denote by $\epsilon_i$ and $\epsilon_i^\prime$ the integers such 
that $s^{(i)}_1=3s^{(i-1)}_1+s^{(i-1)}_2-\epsilon_in$, and 
$s^{(i)}_2=s^{(i)}_1+s^{(i-1)}_2+33-\epsilon^\prime_in$. Since $0\le s^{(i)}_1,s^{(i)}_2<n$,
it follows that $0\le\epsilon_i\le 3$ and $0\le\epsilon^\prime_i\le 2$ and 
$s^{(1)}_2=s^{(1)}_1+s^{(i-1)}_2+33-\epsilon^\prime_in$.

Fix a user ${\cal U}$ with hashed password $p$. Denote by $F(p,c)$ the function
that from a (hashed) password $P$ and a (hashed) challenge $c$ produces the
checksum $w$ as explained in the previous section. Suppose that a pair $(c,w)$
of challenge and response corresponding to user ${\cal U}$ is known (e.g., to the
attacker).  Denote, as in the previous section, the (32) more significant bits of $p$
and $c$ by $p_1$ and $c_1$, and the (32) least significant bits by $p_2,c_2$ 
respectively. 

\begin{remark}
\label{w9}
By applying the algorithmic procedure just described to $w[k]=(w_1\xor k\,\|\,
\ldots\,\|\, w_8\xor k)$ for $0\le k<32$ only one value of $k$ produces a 
nonempty output, hence that value of $k$ is precisely $w_9$, i.e. we have 
$w_9=k$. 
\end{remark}
We will not prove this remark, since this result is totally dependent
on the specific parameters used for this authentication scheme.  However, 
we do give a mild justification. Suppose that we have made our choice for
a $w_9$ candidate, say $\hat w_9$, and that it is a wrong choice. Suppose 
furthermore that the $\delta_i$ are already chosen. Then the polygon ${\cal P}$ 
defined by this choices  is
\begin{align*}
{\cal P}=\displaystyle\bigcap_{1\le i\le 8} \Big\{(x,y)\in\R^2: &
\frac{n}{31}(w_i\xor w_9\xor\hat w_9 -64)+\delta_in \le \\ &
\alpha_i\cdot X+\beta_iY +\gamma_i\cdot 33 < \\ &
\frac{n}{31}((w_i\xor w_9\xor\hat w_9-63)+\delta_in
\Big\}.
\end{align*}

Notice that each of the sets defined between brackets appearing in 
the above intersection is a polygon, furthermore the two underlying 
lines at each intersection 
$\frac{n}{31}(w_i\xor w_9\xor\hat w_9 -64)+\delta_in =
\alpha_i\cdot X+\beta_iY +\gamma_i\cdot 33 $ and
$\frac{n}{31}(w_i\xor w_9\xor\hat w_9 -63)+\delta_in =
\alpha_i\cdot X+\beta_iY +\gamma_i\cdot 33 $
are at a vertical distance of an integer factor of $\frac{n}{31}$.  When 
the equations come out {\em incorrectly} the polygon they define is 
shifted vertically (upwards or downwards) a distance which is a factor 
of $\frac{n}{31}$. This last fact, together with the fact explained in the 
previous remark that the tangents of the lines of each of these polygons
are quite similar implies our assertion. 

In Figure 2 we can see an image of the result of Procedure 1, four {\em rows}
of twelve polygons each. As we explained, this is a typical behavior.

\begin{figure}
\centering
\setlength\fboxsep{0pt}
\setlength\fboxrule{1pt}
\fbox{\includegraphics[width = 7cm]{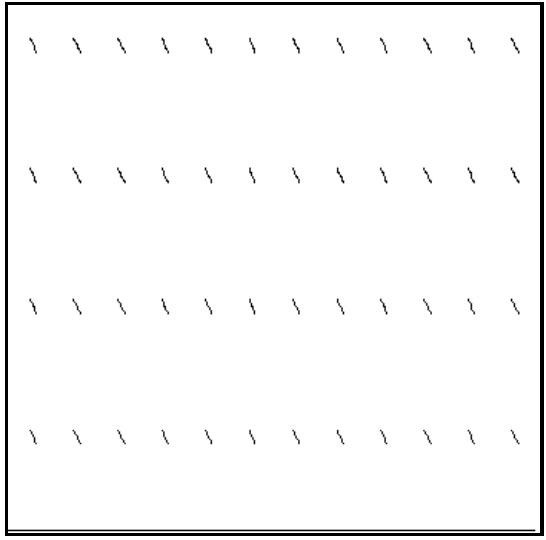}}
\caption{The polygons collection ${\cal P}$}
\label{variospoligonos}
\end{figure}

\end{subsection}
\begin{subsection}{Wash out of invalid passwords}
Let be given a collection $(c,w,{\cal P}),(c^\prime,w^\prime,{\cal P}^\prime)$ 
and an integer $k$, $1\le k\le 32$, such that $(c,w)$ and $(c^\prime,w^\prime)$
are two pairs of challenge and response, and that ${\cal P}$ and ${\cal P}^\prime$
are the set of polygons respective to the given pairs of challenge and response,
and as produced with the Procedure 1.

To describe the output we define some notation. For every $0\le i,j< 2^k$, let
$$
Q\quad:=\quad\left[i\cdot 2^{32-k},(i+1)\cdot^{32-k}\right]\times
\left[j\cdot 2^{32-k},(j+1)\cdot 2^{32-k}\right].
$$ 
And let ${\cal Q}$ denote the set of all these squares. Then, let ${\cal P}_1$
denote the set ${\cal P}_1:=\cup\,(P\cap Q)$, where the union is taken over all
$Q\in{\cal Q},P\in{\cal P}$. Then the output is the subset of ${\cal P}_1$ of
all $P_1\in{\cal P}_1$  for which there exists a $P^\prime\in{\cal P}^\prime$
such that 
$$
(P\xor c)\cap (P^\prime\xor c^\prime)\neq\emptyset ,
$$
this means that there exists a point $(a_1\xor c,b_1\xor c)=(a^\prime\xor c^\prime,
b^\prime\xor c^\prime)$ for some $(a,b)\in P$ and $(a^\prime,b^\prime)\in 
P^\prime$ (so that $f(a,b)=s$ and $f(a^\prime,b^\prime)=s^\prime$). 

The output is a collection of polygons $\widetilde{\cal P}$ 
consisting of all the polygons $\widetilde P=\bigcup\;(P\cap Q)$, the union 
taken over all the squares $Q$ of side $2^{32-k}$ and vertices with entries in 
$\{i\cdot 2^k:0\le i\le 2^{32-k}\}$, and all the polygons $P\in{\cal P}$ in ${\cal P}$ 
for which there exists a $P^\prime\in{\cal P}^\prime$ such that 
$(P\xor c)\cap (P^\prime\xor c^\prime)\neq\emptyset$.   In Figure 2, we see 
two steps of this procedure over the chosen example.

\begin{figure}
\centering
\includegraphics[width = 13cm]{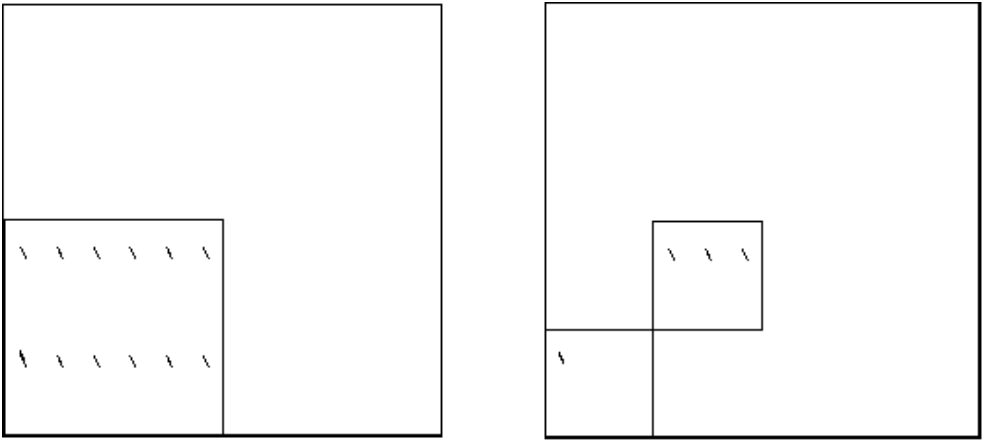}
\caption{The second and third steps}
\label{deados}
\end{figure}

Notice that
we do not calculate $\displaystyle\cup_{P\in{\cal P},P^\prime\in{\cal P}^\prime}
P\cap  P^\prime$, this would be inefficient, e.g. it would make the size of the 
output grow and would not make much of a difference in the point size of the 
output.

\end{subsection}
\begin{subsection}{Rising of passwords}
Procedure 3 is straight forward and needs not much explanation. Given a set of
points and a pair of challenge and response, ones simply browses over every 
point of this set calculating the checksum corresponding to this point and the
given challenge and adds it to the output set only if it produces the given response.

\end{subsection}
\begin{subsection}{The complete algorithmic attack}
A complete attack to a valid user, who has produced the pairs of challenge
and response $(c_1,w_1),\ldots,(c_t,w_t)$ is done by repeated application of 
the procedures we have just described. To start with, we select the number 
$k\le t$ of these challenge and response pairs to which we are going to apply 
to Procedure~1. That is for the pairs $(c_1,w_1),\ldots,(c_k,w_k)$ we apply 
the Procedure 1and output collections ${\cal P}_1,\ldots,{\cal P}_k$. Typically 
---on our examples--- the number $k$ is taken to be 5 or less. 

On a second step, after selecting a second integer $1\le k^\prime\le 32$ , we
apply Procedure~2 recursively to the triplets $(c^{(1)},w^{(1)},{\cal P}^{(1)}),
\ldots ,(c^{(k)},w^{(k)},{\cal P}^{(k)})$; that is first we apply Procedure 2 to 
$(c^{(1)}1,w^{(1)},{\cal P}^{(1)})$  and $(c^{(2)},w^{(2)},{\cal P}^{(2)})$ using
cubes of size $2^{32-k^\prime}$, then we apply Procedure~2 to the previous 
result and $(c^{(3)},w^{(3)},{\cal P}^{(3)})$, and continue this recursive 
application until the $k$--th tuple is reached. After doing this (the $k$ 
applications of this procedure) we get a set of polygons $\widetilde{\cal P}$ 
having a small number of integer points (compared to the brute--force value 
$2^{64}$).

Finally, we recursively apply Procedure~3 to the set $\widetilde {\cal P}$ and
each of the remaining challenge and response  pairs $(c^{(k+1)},w^{(k+1)}),
\ldots,(c^{(t)},w^{(t)})$ as follows. First we extract every integer point of  
$\widetilde {\cal P}$ and store them as points. Then we use Procedure~3 
with the resulting set as input and $(c^{(k+1)},w^{(k+1)})$. After we have 
finished with $(c^{(k+j)},w^{(k+j)})$, for $j\ge 1$, we continue by applying
Procedure~3 to the resulting set and $(c^{(k+j+1)},w^{(k+j+1)})$. We end 
when we $(c^{(t)},w^{(t)})$ is reached and there are no more challenge and 
response pairs left, or before if the set of remaining points has only one 
point left. In the latter case we have found the password's hash. Else, we 
get as output the set of passwords' hashes that pass every one of the 
challenge and response pairs we have as input. It should be remarked,
and shall be furtherly emphasized by empiric data in the next section, 
that in the case of the output being a set of more than one point, all of these 
points are not just random points in $[0,2^{32}]\times [0,2^{32}]$, but have 
high probability of passing an additional test.

\end{subsection}

\end{section}

\begin{section}{Statistics and Conclusions}
\label{stats}
We coded the algorithm of the preceding section in Squeak Smalltalk, and
analyzed the results. 
In the examples tested, about 300 possible passwords were left with the
use of only 10 pairs of challenge and response. Notice that in a plain
brute--force attack about $2^{64}-300=18,446,744,073,709,551,316$ would 
remain as possible passwords. It took about $100$ pairs of challenge and 
response to cut the $300$ points set to a set containing 2 possible passwords
(i.e., a fake passwords and the password indeed). Finally it took about $300$
pairs of challenge and response to get the password.

In other examples we used only ten pairs of challenge and response, 
getting thus a set of approximately 300 points in each case. Then we 
randomly selected 1000 challenges and made the 1000 tests to every
one of the 300 points we had. The result was that each of the remaining 
points passed over 920 of the 1000 tests. That is, the mean (sample--mean)
of the probability a point (in the set left after applying our algorithm to
ten pairs of challenge and response) is of 92\%.

We therefore are able to make a variety of attacks depending on the amount
of pairs of challenge and response we get from the user we want to impersonate.
The two extremal cases being very few pairs of challenge and response from
the same user, and a lot of pairs of challenge and response. The second attack,
that of many pairs of challenge and response captured, is straight--forward:
apply the algorithm described above until the password is found. The first
case, that of only a few pairs of challenge and response captured, is as
well easy to carry: simply apply the algorithm we described with all the pairs 
of challenge and response captured, then use any possible password in the 
set produced by the application of the algorithm for authenticating yourself as
a user (some of these fake passwords will still pass many tests!).

We do not analyze the order of the complexities of our procedures because we
are working with an input of fixed size. Since every one of the attacks we made 
was much alike in the performance/computation--time aspect, we describe a 
specific example of an attack pointing out the computation time needed at each 
step in the attack, and the amount of points left (in the $2^{64}$ to $1$ countdown of
possible hashed passwords).
In a specific example we applied Procedure 1 to a pair of challenge and response
calculating $48$ polygons, each a traslation of the other, of area $2^{33}$, which
makes a set of area approximately $2^{38}$. The whole procedure application lasted
no more than half an hour in a Pentium 3 64Mb RAM personal computer. We
applied Procedure 1 to four other pairs of challenge and response obtaining
similar results. This constituted the first two hours and a half of the attack.
The four applications of Procedure 2 lasted half an hour each, and resulted in a
collection of 32 polygons of area approximately $2^{16}$ each, i.e. a set of
size $2^{21}$. Finally we applied Procedure 3 getting the announced 300 possible
hashed passwords in about six hours and using only 10 pairs of challenge and 
response. The complete attack lasted approximately twelve hours, in this case
we had available some 300 more challenge and response pairs and were able
to recover the hashed password in five more minutes.

\end{section}


\vskip 1cm
\overfullrule 0pt
\hrule 

\bigskip

\noindent{\bf Credits:}
This vulnerabilities were found and researched by Agustin Azubel, 
Emiliano Kargieman, Gerardo Richarte, Carlos Sarraute and Ariel 
Waissbein of CORE Security Technologies, Buenos Aires, Argentina. 

A prior notification of these results appeared signed by researchers
and co--author Ivan Arce as ``An advisory on MySQL's login protocol" 
in SecurityFocus' Bugtraq newsgroup \cite{core}. 

\vskip 0.5cm
\overfullrule 0pt
\hrule 

\bigskip



\end{document}